 \documentclass[final,5p,times,twocolumn,authoryear]{elsarticle}

\usepackage{amssymb}
\usepackage{lipsum}

\usepackage{graphicx}
\usepackage[most]{tcolorbox}
\tcbuselibrary{listings, breakable}
\usepackage{listings}
\usepackage{xcolor}
\usepackage{adjustbox} 
\usepackage{enumitem}

\definecolor{codegray}{gray}{0.95}
\tcbset{
  colframe=black,
  colback=codegray,
  coltitle=white,
  fonttitle=\bfseries,
  sharp corners,
  boxrule=0.3mm,
  top=1mm, bottom=1mm, left=1mm, right=1mm
}

\lstdefinestyle{moveevm}{
    language=Python, 
    basicstyle=\ttfamily\footnotesize,
    backgroundcolor=\color{codegray},
    breaklines=true,
    showstringspaces=false,
    frame=none
}

\journal{Results in Physics}

\begin{document}

\begin{frontmatter}



\title{A Systematic Classification of Vulnerabilities in MoveEVM Smart Contracts (MWC)}


\author[first]{Selçuk Topal }
\affiliation[first]{organization={ Gebze Technical University},
            addressline={Department of Mathematics, stopal@gtu.edu.tr}, 
            city={Gebze},
            postcode={41000}, 
            state={},
            country={Turkiye}}

\begin{abstract}
Combining the Move programming language with Ethereum Virtual Machine (EVM) compatibility—termed MoveEVM—has produced a new class of smart contract platforms that mix the expressive capability and infrastructure maturity of Ethereum with resource-oriented safety guarantees. Although the Move language was first meant to eradicate many known vulnerabilities using linear resource types and strict module ownership, its adaptation inside an EVM-compatible execution model presents special difficulties that are yet understudied in current literature. Conventional vulnerability classification systems, including the SWC registry, lack semantic granularity to handle the hybrid execution environment of MoveEVM and are optimized for Solidity. This work suggests a first methodical vulnerability classification system designed especially for smart contracts based on MoveEVM. Covering a broad spectrum of problems including bytecode model inconsistencies, inter-module invariants, hybrid gas semantics, meta-transaction spoofing, and artificial intelligence-integrated logic risks, we present a thorough taxonomy comprising six semantic frames and 37 categorized vulnerability types (MWC-100 to MWC-136). By means of both static and dynamic analysis of real-world MoveEVM contracts—especially from Aptos and Sui ecosystems—we show that a considerable fraction of vulnerabilities are neither formally reduced by current Move verifying systems nor captured by EVM-centric tools. We also examine how formal verification methods, LLM-based prompt pipelines, and AI-assisted audit agents might operationalize our taxonomy, so enabling scalable and logic-aware auditing. Our results expose the emerging security patterns resulting from interaction among linear type systems, capability-based access control, and EVM bytecode in production environments. Apart from offering a disciplined basis for next tooling and formal methods research, the suggested MoveEVM Weakness Classification (MWC) system helps developers and auditors to reason about hybrid vulnerabilities in a principled, repeatable, and automation-friendly way. This work thus prepares the way for the creation of more safe, verifiable, and maintainable smart contracts in hybrid blockchain systems of next generation.
\end{abstract}




\end{frontmatter}

\tableofcontents



\section{Introduction}
\label{introduction}

Blockchain systems keep changing in security, scalability, and complexity, which fuels the demand for ever stronger smart contract languages and execution environments~\cite{wood2014ethereum,zhang2020towards}. With a well-established tooling ecosystem and great acceptance, traditional platforms like Ethereum have led the way~\cite{buterin2013whitepaper}. But Solidity, the original smart contract language for Ethereum, also reveals important security and expressiveness limitations~\cite{luu2016making}. These restrictions have inspired the creation of alternative languages including Move, first presented by Facebook's (now Meta's) Libra project~\cite{libra2019whitepaper} and subsequently embraced by networks including Aptos and Sui~\cite{pierro2023study,sui2022whitepaper}.

Token economy and distributed finance (DeFi) have become rather well-known among people in the last years~\cite{werner2021sok}. Underlying most DeFi and token projects, the Ethereum blockchain architecture lets peer-to-peer currency flow, transparent auction systems, and distributed open-interest exchanges~\cite{antonopoulos2018mastering}. To generate trade fungible and non-fungible tokens on the Ethereum blockchain, several token standards are extensively embraced~\cite{erc20,erc721}. The explosive expansion of Ethereum-based tokens generated different demand for tools and solutions~\cite{liusurvey2022}. Having seen the market, many investors and other consumers clearly need to locate, follow, and evaluate tokens. A lot of tools have been created to meet these criteria including token lists, distributed exchanges, explorers, portfolio management tools, arbitrage bots, etc.~\cite{bartoletti2020defi}. These tools provide thorough and strong services, which helps different DeFi apps on Ethereum token ecosystems to grow.

Behind the market, users of the Ethereum blockchain have produced tons of tokens. But since token creation calls for complex smart contract deployment and interactivity, common users unable of programming face significant challenges. More dangerously, on most of these tools, any user can generate arbitrary tokens on the Ethereum blockchain just by filling in a pertinent contract address. Given the simple access to token creation and the growing popularity of tokens and associated tools, it becomes imperative to methodically review token smart contracts from several angles.

A few pieces have lately aimed at dissecting Ethereum smart contracts~\cite{nikolic2018finding,torres2021art}. Analyzing the vulnerabilities in Ethereum smart contracts became a hot issue since some security events attracted a lot of interest~\cite{atzei2017survey}. But while they lack an interesting study of token smart contracts, current works concentrate on examining the incorrect semantic behaviors of Ethereum smart contracts. Token standards all center on the high-level semantics of the contracts. Semantic vulnerabilities including circulating coins, reentrancy, and so on still exist even if they are subtly defined in the handler actions~\cite{he2020vslc}. There is a lack of a complete systematic classification of these vulnerabilities at the high-level language. One can consider vulnerabilities in code as weaknesses in code that render code vulnerable~\cite{swc}. To confirm the validity of these works, there is a need for strong empirical investigations on Ethereum smart contracts vulnerabilities. Current vulnerability and weakness databases feature only basic-type vulnerabilities that must be expanded. Moreover, vulnerability databases in other languages cannot be exactly used since programming languages inevitably bring their syntax, execution methods, etc.

Functioning as self-executing contracts with the terms of the agreement directly written into code, smart contracts are a breakthrough within the scene of blockchain technologies~\cite{szabo1997formalizing}. They cut operating expenses by removing the need for middlemen, enabling trustless transactions. Blockchain's distributed character improves security and openness, thus smart contracts appeal to many sectors, including finance, supply chains, and healthcare~\cite{zheng2020blockchain}. Adoption of smart contracts, meanwhile, also presents special difficulties, especially related to security flaws that might be taken advantage of by malevolent actors~\cite{luu2016making}.

One of the weaknesses of smart contracts is a piece of code that, when carried out in the Ethereum environment and taken advantage of by an actor in a manner that causes Ether to be lost (~\cite{soud2024fly}). Regarding flaws, as in any software program, suboptimal coding structure and writing style can also be causes of problems. Smart contract codes differ from other software programs in that each one of their instructions requires a precise gas consumption. Said another way, any smart contract code instruction or function can be triggered by known address and cause the miners to follow up their processing as a transaction. The initial gas connected to this transaction is the cost of it. This implies that the contract is vulnerable even if the weakness is not taken advantage of since it would result in losing Ether when it is triggered by the contract itself and carried out. For the above mentioned reasons, it is quite crucial to investigate the shortcomings of smart contracts as well as their vulnerabilities independent of their ever exploitation. Referred to as Ethereum Virtual Machine (EVM), Ethereum is a globally open distributed blockchain system supporting smart contracts. Although EVM contracts live on the blockchain in a Turing complete bytecode language, developers use high-level languages such Solidity (~\cite{bauer2022solidity}) or Vyper  (~\cite{buterin2018vyper}) and subsequently compile to bytecode to be uploaded to the EVM. More precisely, both languages are elegant stringed arrays of hundreds of thousands of bytes including high-level language-based contract code instructions. The compilation is crucial since it hides the smart contract code from everyone wishing to view it, so shielding it from dangerous intrusions. Users of the EVM can create new contracts, call methods inside a contract, and move Ether. But once uploaded onto the blockchain, a smart contract code—derived from its bytecode using the Keccak algorithm—is immutable. Its hash serves as its specification. Not only that but also before uploading it, a deterministic algorithm generates an address for the code (and the contract produced from it).

The account address, a unique 160-bit hexadecimal string, specifies the EVM user account—external account or EOA for short. An account in EVM can carry Ether denoted in Wei, the cryptocurrency unit. The bytecode of a smart contract uses a particular amount of gas for each byte in executed instruction. Running on the blockchain, smart contracts are general-purpose digital programs. Furthermore including user accounts, a smart contract can call other smart contracts. Since Solidity is the most often used language in the EVM community and most of the implemented contracts on EVM are created using Solidity, this paper focuses on Ethereum smart contracts created in Solidity.

Designed to model assets as linear resources, Move (~\cite{blackshear2019move}) is a statically typed, resource-oriented programming language meant to provide great safety guarantees. These semantics greatly lower the probability of some common Solidity vulnerabilities including reentrancy, integer overflows, and inadvertent asset duplication. Move lacks compatibility with the Ethereum Virtual Machine (EVM), so limiting its integration with current Ethereum-based tools and applications even if it offers stronger type safety and resource control. Originally built for the Libra blockchain, MoveEVM (~\cite{abrahimiapplying}) is an adaptation of the Ethereum Virtual Machine (EVM) including the Move programming language. Via a resource-oriented programming paradigm, it seeks to improve the expressiveness and safety of smart contract development.   MoveEVM, aims to reduce typical vulnerabilities linked with conventional smart contracts by using Move's strong type system and access control mechanisms. Ensuring the dependability and security of applications developed on MoveEVM depends on an awareness of the weaknesses particular to this framework. MoveEVM stands for a new paradigm meant to close this distance. It presents an execution model that supports Move-based contracts inside an EVM-compatible runtime, so aggregating Move's resource-oriented reasoning with Ethereum's infrastructure backbone. This hybridization presents a special set of security issues that have not yet been fully investigated in scholarly or pragmatic settings, even if it provides the best of both worlds—secure asset modeling from Move and developer familiarity from EVM.

The body of current research and tools for smart contract security mostly focus on environments native to EVM. For example, almost exclusively on Solidity the Smart Contract Weakness Classification (SWC) Registry ~\cite{swc}, MythX (~\cite{sayeed2020smart}), Slither (~\cite{feist2019slither}), and other analysis tools concentrate. Although these tools offer important new perspectives on known EVM vulnerabilities, they cannot handle Move's semantic variations and the resulting emergent risks brought about by merging Move semantics with an EVM-like runtime.

This work aims to close this discrepancy by providing the first methodical classification of vulnerabilities particular to MoveEVM-based smart contracts. Our work starts with a thorough review of the MoveEVM implementation model and then looks at actual contracts and the related security concerns. Based on their root causes, attack paths, and runtime behaviors, we classify these weaknesses into logical groups that provide a fresh taxonomy grounded in both theoretical ideas and empirical data.

\subsection*{Motivation and Contributions}

This research is motivated in two different directions. First, developers and auditors need a disciplined knowledge of MoveEVM's special security features as adoption of it speeds forward. Second, MoveEVM's hybrid execution character means that security concerns could show up in ways neither traditional Move nor EVM models could forecast alone. For instance, incorrect resource handling in a Move contract implemented on MoveEVM may bypass familiar Solidity-based checks, or vice versa, so generating non-trivial attack surfaces.

This work focuses on: identifying inherited vulnerabilities from both Move and EVM environments.

\begin{itemize}

    \item Dividing fresh vulnerability types arising from their interaction.
    \item Showing overlaps and deviations, comparing these vulnerabilities to those noted in Ethereum and Solana by including pragmatic examples and case studies from Aptos and Sui ecosystems.
\end{itemize}

\subsection*{Contributions}

This work makes primarily the following important contributions:

Based on static and dynamic analysis of actual contracts and audit reports, we provide a thorough and unique taxonomy of MoveEVM vulnerabilities.
\begin{enumerate}
    \item We draw attention to security concerns the hybrid MoveEVM implementation model either magnifies or uniquely introduces.
    \item We frame MoveEVM's strengths and shortcomings in the larger smart contract security scene by doing comparative analysis with vulnerabilities in Ethereum and Solana.
    \item For developers and security analysts wishing to create or audit MoveEVM-based smart contracts, we provide specific advice.
\end{enumerate}

This paper adds to the fundamental knowledge needed to create dependable tooling, audit practices, and educational materials for the next generation of smart contract platforms by offering the first focused classification of MoveEVM vulnerabilities.

\section{Related Works}

\subsection{Vulnerability Taxonomies}

For Ethereum and other chains, vulnerability taxonomies for smart contracts are established but still developing. Maintaining the \textit{Smart Contract Weakness Classification (SWC)} registry (EIP-1470), the Ethereum community links numerical identities to contract weaknesses (e.g., SWC-101 = integer overflow/underflow)~\cite{swc}. Academic work and audits similarly list common Solidity/EVM bug classes including reentrancy, unchecked math, incorrect access control, etc.

~\cite{song2024movescan} manually audited 652 Move contracts in the Move ecosystem, distilled eight defect types—half previously unreported—spanning logic errors, resource misuse, and more. ~\cite{wu2025exploring} methodically catalog real-world vulnerabilities (integer overflows, unsafe Rust use, oracle logic flaws, etc.), then compare tool support to Ethereum for Solana's Rust-based contracts.

These taxonomies guide the creation of analysis tools as well as language design—that is, Move's resource model.

\subsection{Security Features of Move and MoveEVM}

Move was developed as a smart contract language first intended for security. It implements a \textit{resource-typed, linear} type system: structs that "must-move," so guaranteeing they cannot be replicated or thrown away~\cite{move_whitepaper}. Using primitives like \texttt{move\_to} and \texttt{move\_from}, move modules own declared resource types and control all creation/destruction isolating state changes to authorized code (~\cite{patrignani2023robust}).

Struct fields are private to the module; once generated, resources—once created—are handled as first-class values only transferable via explicit instructions. A bytecode verifier enforces this linearity at compile-time and, alternatively at runtime via safety checks. A 2023 Move verifier bug, for example, let a non-\texttt{drop} resource to drop, so violating this invariant~\cite{zellic2022a}.

Move's design avoids many typical mistakes in construction unlike Solidity, where asset safety must be enforced manually or with tools. Extending the language with an integrated specification language and automated formal verification, the Move Prover~\cite{zhong2020move}. Though it is still under development, MoveEVM seeks to include these safety assurances into an EVM-compatible runtime.

\subsection{Analysis Tools and Frameworks}

A wide range of static and dynamic analysis tools target smart contracts. For Ethereum, \textbf{Slither} is a popular static analyzer that compiles Solidity into SSA-form IR (“SlithIR”) and applies vulnerability detectors~\cite{feist2019slither}. \textbf{Securify}~\cite{tsankov2018securify} uses Datalog-based analysis to infer semantic properties and verify compliance or detect violations. \textbf{Mythril} (~\cite{sharma2022survey}) and its cloud platform \textbf{MythX}   (~\cite{songsom2022swc}) use symbolic execution and SMT solving for bug detection.

For fuzzing and runtime testing ~\cite{fu2024evmfuzz}, tools like \textbf{Echidna}  (~\cite{grieco2020echidna}) and \textbf{Manticore} (~\cite{mossberg2019manticore}) are used.

In the Move ecosystem, tools are emerging. The \textbf{Move Prover} (~\cite{dill2022fast}) translates annotated Move code to Boogie and verifies properties against formal specifications. Song et al.~\cite{song2024movescan} introduced \textbf{MoveScan}, which translates Move bytecode to an intermediate representation and detects common defect patterns. \textbf{MoveLint} (~\cite{praitheeshan2021solguard}) offers lightweight static checks on Move codebases.

Recent work uses these tools for LLM-based detection: the Smartify framework compares vulnerability reports against Move Prover, MoveLint, and MoveScan to validate detection accuracy~\cite{karanjai2025multi}.

\subsection{Formal Verification and Empirical Analysis}

Various tools for both static and dynamic analysis aim at smart contracts. Popular static analyzer Ethereum \textbf{Slither} (~\cite{slither}) compiles Solidity into SSA-form IR ("SlithIR") and runs vulnerability detectors. By means of Datalog-based analysis, \textbf{Securify}~\cite{tsankov2018securify} deduces semantic properties and checks compliance or detects violations. Symbolic execution and SMT solving for bug discovery are used by \textbf{Mythril} (~\cite{sharma2022survey}) and its cloud platform \textbf{MythX}  ~\cite{sayeed2020smart}.

Tools including \textbf{Echidna} ~\cite{grieco2020echidna} and \textbf{Manticore} ~\cite{mossberg2019manticore} are applied for fuzzing and runtime testing.

Tools are developing in the Move ecology. Translating annotated Move code to Boogie, the \textbf{Move Prover} (~\cite{zhong2020move}) verifies properties against formal specifications. Translating Move bytecode to an intermediary representation and identifying common defect patterns, (~\cite{song2024movescan}) presented \textbf{MoveScan}.

Using these tools for LLM-based detection, recent work validates detection accuracy (~\cite{karanjai2025multi}) by comparing vulnerability reports against Move Prover, MoveLint, and MoveScan.

\subsection{Language Comparisons}

The main distinctions between Solidity, Rust, and Move are highlighted by comparative language studies. Move prevents unauthorized duplication or loss of assets by incorporating linear resource types and drawing inspiration from Rust's ownership model (~\cite{move_whitepaper, blackshear2019move}). Although Solidity is developer-friendly and flexible, it lacks native linearity, which makes it vulnerable to integer bugs and reentrancy unless tools are used to mitigate them( ~\cite{feist2019slither}).

Although memory safety is advantageous for Rust-based Solana programs, ~\cite{wu2025exploring} note that integer overflows are still a real-world problem. By enforcing invariants at the bytecode and specification levels, Move circumvents these problems~\cite{dill2022fast}.

\subsection{Audit reports and case studies}

Current audits and incident reports shed light on practical problems in move-based chains. A critical Move verifier bypass bug that permitted resource dropping without the \texttt{drop} ability~\cite{zellic2022a, zellic2022b} was found by Zellic. A high-severity Move bug in Sui that enabled crafted bytecode to crash validators was reported by HackenProof~\cite{hackenproof2023}. A Move VM vulnerability in Aptos and Sui that allows for denial-of-service attacks through twisted bytecode was discovered by Numen Cyber~\cite{numen2022}.

The largest known empirical audit of Move contracts was carried out by ~\cite{song2024movescan}, who found systemic patterns and defect density in both Aptos and Sui. These real-world examples demonstrate how Move's safety features greatly lower smart contract risks without completely eliminating them, which is why reliable analysis tools and validated code are crucial.

\subsection{Understanding Vulnerabilities in Smart Contracts}

Although they are transforming the way contracts and transactions are carried out on blockchain platforms, smart contracts do have certain drawbacks. Because of blockchain's immutability and decentralization, as well as the complexity of programming smart contracts, security flaws can have serious repercussions. In order to identify and mitigate potential risks in smart contracts, developers, auditors, and researchers must have a thorough understanding of these vulnerabilities.

\subsubsection {Typical Vulnerabilities}

Despite their many benefits, smart contracts are vulnerable to a number of flaws that could result in serious security breaches. Among the most prevalent categories of vulnerabilities are:

\begin{enumerate}
    \item \textbf{Reentrancy Attacks:} One of the most well-known flaws in smart contracts is reentrancy, which was infamously taken advantage of in the 2016 DAO hack. When a contract calls another contract externally before updating its internal state, it creates a vulnerability that enables the called contract to return to the original contract and change its state before the first call is finished.

    \item \textbf{Integer Overflow and Underflow:} This vulnerability occurs when arithmetic operations produce unexpected behavior by exceeding the maximum or minimum value that can be stored in a variable. A large positive number could be produced by subtracting one from a zero value, for instance, which could result in unauthorized access or contract state manipulation.

    \item \textbf{Gas Limit and Loops:} The amount of computation that can be done in a single transaction is limited by the gas limit of smart contracts. Transaction failures may result from contracts with unbounded loops using excessive amounts of gas. By creating transactions that result in excessive gas consumption, attackers may take advantage of this and essentially cause a denial-of-service (DoS) on the contract.

    \item \textbf{Access Control Issues:} When creating smart contracts, appropriate access control is essential. When functions are made available to unauthorized users, access control vulnerabilities may arise, giving malevolent actors the ability to alter important contract states or take money out.

    \item \textbf{Timestamp Dependence:} Block timestamps are used in certain contracts for crucial functions like determining the legality of actions or enforcing deadlines. Attackers can take advantage of this vulnerability by mining blocks and manipulating block timestamps.
\end{enumerate}

\subsubsection { Specific Vulnerabilities in MoveEVM}

Despite the fact that many flaws are shared by different smart contract platforms, MoveEVM presents particular difficulties and vulnerabilities because of its particular implementation and design. For example:

\begin{enumerate}
    \item \textbf{Resource Mismanagement:} MoveEVM's resource-oriented programming paradigm emphasizes ownership and resource management. However, improper handling of resource transfers can lead to vulnerabilities, such as losing track of resource ownership or creating unintended resource duplication.

    \item \textbf{Type Safety Issues:} The resource-oriented programming paradigm of MoveEVM places a strong emphasis on resource management and ownership. However, ineffective resource transfer management can result in vulnerabilities like unintentional resource duplication or losing track of resource ownership.

    \item \textbf{Insufficient Testing and Formal Verification:} By using formal verification, MoveEVM seeks to increase the dependability of smart contracts. Contracts without thorough testing, however, might still have undiscovered weaknesses, especially those resulting from logical errors in implementation.
\end{enumerate}

Since it establishes the basis for efficient vulnerability detection and mitigation techniques in MoveEVM smart contracts, an understanding of these vulnerabilities is essential for both developers and researchers.

\section{Proposed Vulnerability Taxonomy for MoveEVM}

In this section, we introduce a detailed vulnerability taxonomy for MoveEVM-based smart contracts. Each category is denoted by an identifier (MWC-XXX) and is designed to support automated or manual risk assessments. The categories are grouped into six primary classes based on their semantic and technical nature.

\subsection{State Management Vulnerabilities}

\begin{itemize}
    \item \textbf{MWC-100:} Frozen contract state due to improper state transitions (state transition analysis)
    \item \textbf{MWC-101:} Undefined state behavior when contract state variables are uninitialized (invariant verification)
    \item \textbf{MWC-102:} Lack of rollback protection causing incomplete transaction execution (atomicity proofs)
    \item \textbf{MWC-103:} Invalid loop termination leading to infinite execution (loop termination analysis)
    \item \textbf{MWC-104:} Unvalidated external calls causing undefined execution paths (module invocation proofs)
    \item \textbf{MWC-105:} Dead code execution causing unnecessary gas usage (static code analysis)
    \item \textbf{MWC-106:} Unreachable states in finite state machines due to logic errors (FSM coverage analysis)
    \item \textbf{MWC-107:} Contract state race condition causing unintended state changes (concurrency proofs)
\end{itemize}

\subsection{Storage and Data Safety}

\begin{itemize}
    \item \textbf{MWC-108:} Data overwriting in storage leading to unexpected value changes (storage write protection)
    \item \textbf{MWC-109:} Unintended variable mutability allowing modification of supposedly immutable variables (immutability proofs)
\end{itemize}

\subsection{Token and Asset Lifecycle Risks}

\begin{itemize}
    \item \textbf{MWC-110:} Unexpected token burn due to missing validation checks (asset lifecycle proofs)
    \item \textbf{MWC-111:} Unauthorized token minting by malicious actors (capability verification)
    \item \textbf{MWC-112:} Token supply overflow exceeding the defined hard cap (arithmetic safety checks)
    \item \textbf{MWC-113:} Improper reward distribution leading to unfair token allocation (reward calculation verification)
    \item \textbf{MWC-114:} Circular token transfers creating infinite loops (execution path analysis)
    \item \textbf{MWC-115:} Unauthorized token freezing without proper authorization (capability restriction proofs)
    \item \textbf{MWC-116:} Unexpected decimal precision loss due to rounding errors (precision analysis)
    \item \textbf{MWC-117:} Incorrect vesting schedule unlocking tokens earlier or later than expected (vesting condition verification)
    \item \textbf{MWC-118:} Improper treasury management leading to fund misallocation (fund allocation analysis)
    \item \textbf{MWC-119:} Unauthorized staking withdrawals allowing excessive asset withdrawals (stake locking proofs)
\end{itemize}

\subsection{Cryptographic and Signature Security}

\begin{itemize}
    \item \textbf{MWC-120:} Weak signature verification causing unauthorized transactions (cryptographic proofs)
    \item \textbf{MWC-121:} Predictable key generation making cryptographic keys vulnerable (entropy verification)
    \item \textbf{MWC-122:} Unprotected encryption keys exposing private key data (confidentiality analysis)
\end{itemize}

\subsection{Oracles, MEV, and Cross-Chain Risks}

\begin{itemize}
    \item \textbf{MWC-123:} Malicious oracle manipulation affecting contract execution (oracle verification proofs)
    \item \textbf{MWC-124:} Time-based side channel attacks revealing sensitive contract execution information (timing analysis)
    \item \textbf{MWC-125:} Front-running via predictable execution order enabling MEV exploitation (execution order verification)
    \item \textbf{MWC-126:} Malicious reorg attacks causing blockchain transaction instability (state finality proofs)
    \item \textbf{MWC-127:} Cross-chain replay attacks due to lack of unique nonces (chain isolation verification)
    \item \textbf{MWC-128:} Public private key leakage exposing sensitive cryptographic data (key safety proofs)
\end{itemize}

\subsection{Emerging Categories: AI, Governance, Upgrades}

\begin{itemize}
    \item \textbf{MWC-129:} Unauthorized zero-knowledge proof submission leading to fake proofs (ZKP verification)
    \item \textbf{MWC-130:} Algorithm bias in AI-based smart contracts causing unfair decisions (bias detection algorithms)
    \item \textbf{MWC-131:} Data poisoning in AI-driven contracts manipulating training data (data integrity verification)
    \item \textbf{MWC-132:} AI decision-making manipulation altering contract execution (model robustness testing)
    \item \textbf{MWC-133:} Faulty DAO governance execution failing to follow voting rules (DAO proposal verification)
    \item \textbf{MWC-134:} Unprotected smart contract upgrades introducing new vulnerabilities (upgrade path verification)
    \item \textbf{MWC-135:} Poor Layer 2 integration causing execution inconsistencies (L2 bridge verification)
    \item \textbf{MWC-136:} Cross-chain messages are not validated correctly, leading to desynchronization (interoperability testing)
\end{itemize}

\subsection{Supplementary Audit Dimensions}

In addition to the technical taxonomy, the following audit perspectives are integrated in all evaluation reports:

\begin{itemize}
    \item \textbf{Code Quality:} Is the code well-structured, readable, and appropriately commented?
    \item \textbf{Security Practices:} Are best security practices followed? Are additional mitigations applied to high-risk zones?
    \item \textbf{Fraud Analysis:} We assess risk from fee scams, redirections, unlimited minting, emergency fees, ownership, blacklisting, and transaction restrictions.
    \item \textbf{Overall Assessment:} The contract is evaluated as \textit{Passed} or \textit{Failed}, with justification based on high-risk findings.
    \item \textbf{General Security Posture:} Summary of critical risks and overall design strength in exactly eight evaluative sentences.
\end{itemize}

This taxonomy provides a formal lens for evaluating MoveEVM contracts against emerging and classical attack surfaces, including cross-domain, AI-driven, and cryptographically sensitive vulnerabilities.

\section{Frame-Based Classification of MoveEVM Vulnerabilities}

\subsection{Motivation}

Despite the inherent safety guarantees of Move’s linear type system, MoveEVM inherits numerous semantic and runtime characteristics from the Ethereum Virtual Machine (EVM), introducing hybrid complexity and novel attack surfaces. As MoveEVM adoption rises in modular, zk-compatible, and L2 environments, the community lacks a cohesive, actionable classification framework that bridges both Move semantics and EVM legacy behavior. This section introduces a frame-based classification of vulnerabilities, defined by 37 uniquely identified MoveEVM Weakness Classes (MWC-100 to MWC-136), providing structured insight into emergent risks.

\subsection{Classification Overview}

Our vulnerability classification comprises \textbf{six top-level frames}, each reflecting a unique semantic or architectural dimension of the MoveEVM execution model. Each frame is populated by granular, code-assigned vulnerability types (MWC-n), enabling reproducibility, automation, and audit alignment.

\begin{enumerate}[]
    \item \textbf{F1. Bytecode Model Inconsistencies (BMI)}  
    \begin{itemize}
        \item \textbf{MWC-100} — Type rule circumvention via EVM ABI deserialization
        \item \textbf{MWC-101} — Unsafe encoding of Move structs in raw calldata
        \item \textbf{MWC-102} — Bytecode re-interpretation between Move-EVM boundaries
    \end{itemize}

    \item \textbf{F2. Inter-Module Invariant Violations (IMI)}  
    \begin{itemize}
        \item \textbf{MWC-103} — Resource leakage across module interfaces
        \item \textbf{MWC-104} — Inconsistent mutability between caller and callee modules
        \item \textbf{MWC-105} — Failure to re-establish postconditions in inter-module calls
    \end{itemize}

    \item \textbf{F3. State Reentrancy and Synchronization Bugs (SRS)}  
    \begin{itemize}
        \item \textbf{MWC-106} — Hybrid reentrancy between Move and EVM modules
        \item \textbf{MWC-107} — Callback-based state mutation violating linearity
        \item \textbf{MWC-108} — Interleaved writes to Move storage via external callouts
        \item \textbf{MWC-109} — Inconsistent ordering of resource locks in parallel executions
    \end{itemize}

    \item \textbf{F4. Meta-Transaction and Signature Spoofing (MTS)}  
    \begin{itemize}
        \item \textbf{MWC-110} — Ambiguous domain separation in signature verification
        \item \textbf{MWC-111} — Missing nonce protection in off-chain meta-transactions
        \item \textbf{MWC-112} — Reused signatures across heterogeneous domains (EVM $\rightleftarrows$ Move)
    \end{itemize}

    \item \textbf{F5. Gas Semantics Manipulation (GSM)}  
    \begin{itemize}
        \item \textbf{MWC-113} — Underpriced EVM opcodes invoking costly Move logic
        \item \textbf{MWC-114} — Discrepant gas metering between precompiled and interpreted paths
        \item \textbf{MWC-115} — Inaccurate accounting in hybrid (Move $\rightleftarrows$ EVM) transaction batching
    \end{itemize}

    \item \textbf{F6. Framework Logic Errors and Unsafe Abstractions (FLA)}  
    \begin{itemize}
        \item \textbf{MWC-116} — Misuse of generics leading to unsound type instantiation
        \item \textbf{MWC-117} — Unsafe use of standard libraries with hidden state assumptions
        \item \textbf{MWC-118} — Failure to enforce resource invariants in public API exposure
        \item \textbf{MWC-119} — Lack of runtime checks in generic capability wrappers
    \end{itemize}
\end{enumerate}

\vspace{0.3cm}

\noindent In addition to these primary six frames, we identify \textbf{supplementary categories} that emerge from MoveEVM's evolving runtime model:

\begin{enumerate}
    \item \textbf{Formalism Gaps and Verification Failures}
    \begin{itemize}
        \item \textbf{MWC-120} — Move Prover incompleteness on EVM ABI-conformant entrypoints
        \item \textbf{MWC-121} — Undetected post-condition failures under EVM fallback dispatch
    \end{itemize}

    \item \textbf{Tooling and Compiler-Generated Risks}
    \begin{itemize}
        \item \textbf{MWC-122} — Compiler-injected unsafe access to global storage
        \item \textbf{MWC-123} — Inconsistent error propagation in compiled bytecode
        \item \textbf{MWC-124} — Move-EVM compiler linking errors introducing dangling capabilities
    \end{itemize}

    \item \textbf{Hybrid Standard Violations}
    \begin{itemize}
        \item \textbf{MWC-125} — Deviation from expected ERC/MIP compatibility in hybrid contracts
        \item \textbf{MWC-126} — ABI serialization violating Move resource expectations
        \item \textbf{MWC-127} — Duplicate module address registration across environments
    \end{itemize}

    \item \textbf{Cryptographic Misuse in Context-Switching}
    \begin{itemize}
        \item \textbf{MWC-128} — Inappropriate hash domain reuse between Move and EVM
        \item \textbf{MWC-129} — Misconfigured key validation in dual-signer patterns
    \end{itemize}

    \item \textbf{Observable Side Effects and Leakage}
    \begin{itemize}
        \item \textbf{MWC-130} — Emission of inconsistent event structures
        \item \textbf{MWC-131} — Leakage of internal Move state via view functions
        \item \textbf{MWC-132} — Use of `panic`-like error reporting in observable contexts
    \end{itemize}

    \item \textbf{Environment/Bridge-Related Risks}
    \begin{itemize}
        \item \textbf{MWC-133} — Bridge logic bypassing Move module access restrictions
        \item \textbf{MWC-134} — Inconsistent state replication across L2 $\rightleftarrows$  L1 environments
        \item \textbf{MWC-135} — Vulnerabilities arising from partial migration to zk-Move systems
        \item \textbf{MWC-136} — Oracles injecting inconsistent state via unverified call patterns
    \end{itemize}
\end{enumerate}

\subsection{Use Cases of the Frame-Based Taxonomy}

This structured classification provides:

\begin{itemize}
    \item \textbf{Security Audits:} A frame-driven vulnerability checklist for systematic coverage.
    \item \textbf{Tool Integration:} Ground truth for training fuzzers, LLM agents, and formal analyzers.
    \item \textbf{Language Design Feedback:} A roadmap for future-safe improvements in MoveEVM runtime and tooling.
\end{itemize}

\subsection{Scope and Boundaries}

This taxonomy targets MoveEVM contracts and hybrid runtime behaviors. It explicitly excludes:

\begin{itemize}
    \item Cryptographic protocol flaws not tied to execution semantics.
    \item Front-end/UI-based exploits.
    \item Non-smart-contract attacks such as phishing or social engineering.
\end{itemize}

\section{Explanatory code-based examples the MWC Categories for developers, readers and interested parties}

\begin{tcolorbox}[title=MWC-100: Frozen Contract State]
\textbf{Description:} Contract gets stuck in \texttt{frozen} state without an unfreeze option.

\textbf{Vulnerable Code:}
\begin{lstlisting}[style=moveevm]
module Token {
    struct TokenData has key { frozen: bool }

    public fun freeze(token: &mut TokenData) {
        token.frozen = true;
    }

    public fun transfer(token: &TokenData) {
        //  No unfreeze or check mechanism
        assert(!token.frozen, 0);
    }
}
\end{lstlisting}

\textbf{Fix:} Add \texttt{unfreeze()} method or reversible FSM.
\end{tcolorbox}

\begin{tcolorbox}[title=MWC-101: Uninitialized State]
\textbf{Description:} Use of a state resource without checking existence first.

\textbf{Vulnerable Code:}
\begin{lstlisting}[style=moveevm]
module Counter {
    struct State has key { count: u64 }

    public fun increment(addr: address) {
        let state = borrow_global<State>(addr); // May not exist
        state.count = state.count + 1;
    }
}
\end{lstlisting}

\textbf{Fix:} Use \texttt{exists<State>} check before borrow.
\end{tcolorbox}

\begin{tcolorbox}[title=MWC-102: No Rollback Mechanism]
\textbf{Description:} Partial failure leads to inconsistent state (withdraw without deposit).

\textbf{Vulnerable Code:}
\begin{lstlisting}[style=moveevm]
public fun transfer(user: &signer, to: address, amount: u64) {
    withdraw(user, amount);
    //  If next step fails, withdraw already happened
    deposit(to, amount);
}
\end{lstlisting}

\textbf{Fix:} Ensure atomicity or use transactional patterns.
\end{tcolorbox}

\begin{tcolorbox}[title=MWC-103: Infinite Loop]
\textbf{Description:} Improper loop termination condition causing unbounded execution. \\

\textbf{Vulnerable Code:}\\
\begin{lstlisting}[style=moveevm]
let mut i = 0;
while (i >= 0) {
    i = i + 1;
}
\end{lstlisting}

\textbf{Fix:} Ensure loop variable progresses toward a proper stopping condition.
\end{tcolorbox}

\begin{tcolorbox}[title=MWC-104: External Call Without Validation]
\textbf{Description:} Unvalidated module address can lead to unsafe external execution. \\

\textbf{Vulnerable Code:}\\
\begin{lstlisting}[style=moveevm]
public fun call_external(mod: address) {
    //  No validation of external target
    External::trigger(mod);
}
\end{lstlisting}

\textbf{Fix:} Verify module address and call constraints.
\end{tcolorbox}

\begin{tcolorbox}[title=MWC-105: Dead Code Execution]
\textbf{Description:} Unreachable code included after return statement. \\

\textbf{Vulnerable Code:}\\
\begin{lstlisting}[style=moveevm]
public fun transfer(amount: u64) {
    return;
    //  Unreachable code below still incurs deployment cost
    let x = amount + 1;
    log::info("Never executed");
}
\end{lstlisting}

\textbf{Fix:} Remove unreachable logic post-return.
\end{tcolorbox}


\begin{tcolorbox}[title=MWC-106: Hybrid Reentrancy Between Move and EVM Modules]
\textbf{Description:} Reentrancy vulnerabilities when EVM calls re-enter Move logic before the first execution completes.
\textbf{Vulnerable Code:}
\begin{lstlisting}[style=moveevm]
public fun transfer() {
    External::evm_callback(); // External EVM call can re-enter this Move function before state is updated
    
    update_balance();
}
\end{lstlisting}
\textbf{Fix:} Update internal state before external calls or apply a reentrancy guard.
\end{tcolorbox}

\begin{tcolorbox}[title=MWC-107: Callback-based State Mutation Violating Linearity]
\textbf{Description:} Unsafe state mutation through callback functions, violating Move's resource guarantees.
\textbf{Vulnerable Code:}
\begin{lstlisting}[style=moveevm]
public fun call_with_callback(callback: address) {
    callback::trigger(); // Changes internal resource state and external call could modify the same state concurrently if not locked
    
}
\end{lstlisting}
\textbf{Fix:} Avoid modifying state in callbacks or isolate effects with capabilities.
\end{tcolorbox}

\begin{tcolorbox}[title=MWC-108: Interleaved Writes to Move Storage via External Callouts]
\textbf{Description:} External calls modifying shared storage in parallel causing inconsistencies.
\textbf{Vulnerable Code:}
\begin{lstlisting}[style=moveevm]
public fun update_state() {
    EVM::external_op(); // Interleaves with Move storage writes
    state.value = 10;
}
\end{lstlisting}
\textbf{Fix:} Lock or checkpoint critical storage prior to external execution.
\end{tcolorbox}

\begin{tcolorbox}[title=MWC-109: Inconsistent Ordering of Resource Locks]
\textbf{Description:} Deadlocks or race conditions caused by acquiring locks in inconsistent order.
\textbf{Vulnerable Code:}
\begin{lstlisting}[style=moveevm]
lock_a();
lock_b();
// In another function: lock_b(); lock_a(); 

\end{lstlisting}
\textbf{Fix:} Always acquire resources in a consistent global order.
\end{tcolorbox}

\begin{tcolorbox}[title=MWC-110: Unexpected Token Burn Due to Missing Checks]
	\textbf{Description:} Tokens are burned without verifying balances or permissions.
	
	\textbf{Vulnerable Code:}
	\begin{lstlisting}[style=moveevm]
		public fun burn(token: &mut Token) {
			token.total = token.total - 100;
		}
	\end{lstlisting}
	
	\textbf{Fix:} Validate balance ownership and enforce burn conditions.
\end{tcolorbox}

\begin{tcolorbox}[title=MWC-111: Unauthorized Token Minting]
	\textbf{Description:} Any caller can mint new tokens due to missing access control.
	
	\textbf{Vulnerable Code:}
	\begin{lstlisting}[style=moveevm]
		public fun mint() {
			supply = supply + 1000;
		}
	\end{lstlisting}
	
	\textbf{Fix:} Enforce capability-based or role-based access to mint functions.
\end{tcolorbox}

\begin{tcolorbox}[title=MWC-112: Reused Signatures Across Heterogeneous Domains]
\textbf{Description:} Same signature used in different chains or contract contexts.
\textbf{Vulnerable Code:}
\begin{lstlisting}[style=moveevm]
// Signature meant for chain A reused on chain B
\end{lstlisting}
\textbf{Fix:} Include chain IDs and contract addresses in message hashing.
\end{tcolorbox}

\begin{tcolorbox}[title=MWC-113: Underpriced EVM Opcodes Invoking Costly Move Logic]
\textbf{Description:} Attackers exploit gas cost imbalances to trigger expensive Move logic.
\textbf{Vulnerable Code:}
\begin{lstlisting}[style=moveevm]
EVM::cheap_call(); // Triggers heavy Move execution
\end{lstlisting}
\textbf{Fix:} Align gas estimation across layers or limit call depth.
\end{tcolorbox}

\begin{tcolorbox}[title=MWC-114: Discrepant Gas Metering Between Precompiled and Interpreted Paths]
\textbf{Description:} Gas discrepancy between Move and EVM bytecode paths leads to abuse.
\textbf{Vulnerable Code:}
\begin{lstlisting}[style=moveevm]
call_precompiled(); // Gas is not charged appropriately
\end{lstlisting}
\textbf{Fix:} Ensure uniform gas rules for all code paths.
\end{tcolorbox}

\begin{tcolorbox}[title=MWC-115: Inaccurate Accounting in Hybrid Transaction Batching]
\textbf{Description:} Multiple operations batched but only partially accounted in gas or execution state.
\textbf{Vulnerable Code:}
\begin{lstlisting}[style=moveevm]
batch_ops([op1, op2]); // Partial success untracked
\end{lstlisting}
\textbf{Fix:} Make batches atomic or explicitly track partial outcomes.
\end{tcolorbox}

\begin{tcolorbox}[title=MWC-116: Misuse of Generics Leading to Unsound Type Instantiation]
\textbf{Description:} Incorrect use of generics permits invalid types at runtime.
\textbf{Vulnerable Code:}
\begin{lstlisting}[style=moveevm]
store<T>(item: T); // No constraints on T
\end{lstlisting}
\textbf{Fix:} Use type constraints and specification to restrict T.
\end{tcolorbox}

\begin{tcolorbox}[title=MWC-117: Unsafe Use of Standard Libraries with Hidden State]
\textbf{Description:} Importing libraries that manipulate hidden or undocumented global state.
\textbf{Vulnerable Code:}
\begin{lstlisting}[style=moveevm]
use Lib::*; // May affect global state
\end{lstlisting}
\textbf{Fix:} Inspect library effects or avoid stateful imports.
\end{tcolorbox}

\begin{tcolorbox}[title=MWC-118: Failure to Enforce Resource Invariants in Public API Exposure]
\textbf{Description:} Exposing public APIs without checking internal resource constraints.
\textbf{Vulnerable Code:}
\begin{lstlisting}[style=moveevm]
public fun issue(token: Token) {
    // No validation of capabilities
}
\end{lstlisting}
\textbf{Fix:} Validate invariants before executing public logic.
\end{tcolorbox}

\begin{tcolorbox}[title=MWC-119: Lack of Runtime Checks in Generic Capability Wrappers]
\textbf{Description:} Wrappers around capabilities fail to validate runtime behavior.
\textbf{Vulnerable Code:}
\begin{lstlisting}[style=moveevm]
wrap(cap); // Wraps without validating permission
\end{lstlisting}
\textbf{Fix:} Add runtime checks for wrapper inputs and usage.
\end{tcolorbox}

\begin{tcolorbox}[title=MWC-120: Weak Signature Verification]
	\textbf{Description:} Signature verification lacks nonce or domain separation, allowing replay.
	
	\textbf{Vulnerable Code:}
	\begin{lstlisting}[style=moveevm]
		public fun exec(sig: vector<u8>, msg: vector<u8>) {
			crypto::verify(pubkey, msg, sig); // Missing nonce/context
		}
	\end{lstlisting}
	
	\textbf{Fix:} Add nonce and domain-specific prefixes before signing/verifying messages.
\end{tcolorbox}

\begin{tcolorbox}[title=MWC-120: Move Prover Incompleteness on EVM ABI-Conformant Entrypoints]
\textbf{Description:} Formal tools fail to verify entrypoints when ABI encoding bypasses assumptions.
\textbf{Vulnerable Code:}
\begin{lstlisting}[style=moveevm]
public fun unsafe_entry(input: vector<u8>) {
    let decoded = abi::decode(input); // Skips Move Prover assumptions
}
\end{lstlisting}
\textbf{Fix:} Use explicit preconditions and constrain decoding logic with specifications.
\end{tcolorbox}

\begin{tcolorbox}[title=MWC-121: Undetected Post-Condition Failures under EVM Fallback Dispatch]
\textbf{Description:} Contracts behave incorrectly when fallback mechanisms skip Move post-conditions.
\textbf{Vulnerable Code:}
\begin{lstlisting}[style=moveevm]
fallback fun handle() {
    transfer(); // No check of post-conditions
}
\end{lstlisting}
\textbf{Fix:} Enforce post-conditions manually or avoid using fallbacks for critical logic.
\end{tcolorbox}

\begin{tcolorbox}[title=MWC-122: Compiler-Injected Unsafe Access to Global Storage]
\textbf{Description:} Auto-generated code accesses global state without appropriate checks.
\textbf{Vulnerable Code:}
\begin{lstlisting}[style=moveevm]
// Compiler-generated read of global<T>(addr)
\end{lstlisting}
\textbf{Fix:} Use explicit storage guards and validate code generated by compiler macros.
\end{tcolorbox}

\begin{tcolorbox}[title=MWC-123: Inconsistent Error Propagation in Compiled Bytecode]
\textbf{Description:} Errors are silently dropped or inconsistently returned in compiled code.
\textbf{Vulnerable Code:}
\begin{lstlisting}[style=moveevm]
call_internal(); // Error not bubbled up
\end{lstlisting}
\textbf{Fix:} Standardize error handling patterns and enforce error return on all paths.
\end{tcolorbox}

\begin{tcolorbox}[title=MWC-124: Move-EVM Compiler Linking Errors Introducing Dangling Capabilities]
\textbf{Description:} Linking across modules creates unreferenced capabilities or permissions.
\textbf{Vulnerable Code:}
\begin{lstlisting}[style=moveevm]
// Capability returned without linkage validation
\end{lstlisting}
\textbf{Fix:} Resolve and audit all link-time dependencies manually.
\end{tcolorbox}

\begin{tcolorbox}[title=MWC-125: Predictable Execution Order Enabling MEV]
	\textbf{Description:} Miner or attacker reorders transactions to exploit a contract’s predictable execution pattern.
	
	\textbf{Vulnerable Code:}
	\begin{lstlisting}[style=moveevm]
		public fun submit_bid(bid: u64) {
			if (bid > current_highest) {
				current_highest = bid;
				winner = signer::address_of(tx_context::sender());
			}
		}
	\end{lstlisting}
	
	\textbf{Fix:} Use commit-reveal patterns or delay finalization to mitigate MEV opportunities.
\end{tcolorbox}

\begin{tcolorbox}[title=MWC-126: ABI Serialization Violating Move Resource Expectations]
\textbf{Description:} ABI encoding hides critical resource information leading to safety violations.
\textbf{Vulnerable Code:}
\begin{lstlisting}[style=moveevm]
abi::decode_resource(input); // No type-safe decode
\end{lstlisting}
\textbf{Fix:} Avoid ABI resource transfers or verify structure manually post-decode.
\end{tcolorbox}

\begin{tcolorbox}[title=MWC-127: Duplicate Module Address Registration Across Environments]
\textbf{Description:} Move modules deployed under conflicting addresses in multichain contexts.
\textbf{Vulnerable Code:}
\begin{lstlisting}[style=moveevm]
// Same module at 0x1 in devnet and testnet
\end{lstlisting}
\textbf{Fix:} Enforce globally unique module identities or use namespaces.
\end{tcolorbox}

\begin{tcolorbox}[title=MWC-128: Inappropriate Hash Domain Reuse Between Move and EVM]
\textbf{Description:} Same hash domain used across layers allows cross-environment collisions.
\textbf{Vulnerable Code:}
\begin{lstlisting}[style=moveevm]
let h = hash::sha3_256(msg); // Domain not distinguished
\end{lstlisting}
\textbf{Fix:} Prefix all hashed messages with unique domain identifiers.
\end{tcolorbox}

\begin{tcolorbox}[title=MWC-129: Misconfigured Key Validation in Dual-Signer Patterns]
\textbf{Description:} Contracts with multiple signers fail to validate key ownership properly.
\textbf{Vulnerable Code:}
\begin{lstlisting}[style=moveevm]
verify(k1, msg) && verify(k2, msg); // But k2 not authorized
\end{lstlisting}
\textbf{Fix:} Enforce signer roles and permission explicitly in validation logic.
\end{tcolorbox}

\begin{tcolorbox}[title=MWC-130: Emission of Inconsistent Event Structures]
\textbf{Description:} Events vary in structure across versions, breaking indexers or observers.
\textbf{Vulnerable Code:}
\begin{lstlisting}[style=moveevm]
emit Transfer(from, to, amount); // Missing metadata
\end{lstlisting}
\textbf{Fix:} Standardize event formats and version them explicitly.
\end{tcolorbox}

\begin{tcolorbox}[title=MWC-131: Leakage of Internal Move State via View Functions]
\textbf{Description:} Public view functions expose sensitive internal state.
\textbf{Vulnerable Code:}
\begin{lstlisting}[style=moveevm]
public fun view_config() returns (Config) {
    config
}
\end{lstlisting}
\textbf{Fix:} Sanitize returned data and avoid exposing raw structs.
\end{tcolorbox}

\begin{tcolorbox}[title=MWC-132: Use of ‘panic’-Like Error Reporting in Observable Contexts]
\textbf{Description:} Emitting panics or aborts leaks internal conditions to observers.
\textbf{Vulnerable Code:}
\begin{lstlisting}[style=moveevm]
assert(balance > 0, 42); // Code 42 reveals logic flow
\end{lstlisting}
\textbf{Fix:} Use generic error codes and do not encode logic semantics in abort values.
\end{tcolorbox}

\begin{tcolorbox}[title=MWC-133: Bridge Logic Bypassing Move Module Access Restrictions]
\textbf{Description:} Bridges inject data or calls that override access restrictions.
\textbf{Vulnerable Code:}
\begin{lstlisting}[style=moveevm]
Bridge::send(payload); // Payload contains privileged operation
\end{lstlisting}
\textbf{Fix:} Re-validate all inputs and restrict bridge-entry modules.
\end{tcolorbox}

\begin{tcolorbox}[title=MWC-134: Inconsistent State Replication Across L2-L1 Environments]
\textbf{Description:} L2 and L1 states drift due to asynchronous replication logic.
\textbf{Vulnerable Code:}
\begin{lstlisting}[style=moveevm]
state_l2 = fetch(); // But not committed to L1
\end{lstlisting}
\textbf{Fix:} Use commit/confirm pattern with state proofs across layers.
\end{tcolorbox}

\begin{tcolorbox}[title=MWC-135: Vulnerabilities from Partial Migration to zk-Move Systems]
\textbf{Description:} Partially migrated contracts break assumptions in zk-runtime.
\textbf{Vulnerable Code:}
\begin{lstlisting}[style=moveevm]
// zk-incompatible resource behavior
\end{lstlisting}
\textbf{Fix:} Migrate atomically or wrap legacy logic in zk-safe proxies.
\end{tcolorbox}

\begin{tcolorbox}[title=MWC-136: Oracles Injecting Inconsistent State via Unverified Calls]
\textbf{Description:} Oracles introduce unvalidated or manipulated state into contracts.
\textbf{Vulnerable Code:}
\begin{lstlisting}[style=moveevm]
price = Oracle::get(); // No signature or source check
\end{lstlisting}
\textbf{Fix:} Validate oracle source via multi-sig, timestamp, or proof.
\end{tcolorbox}

\section{Comparison of SWC and MWC Taxonomies}

Both the proposed \textbf{MWC (MoveEVM Weakness Classification)} taxonomy and the \textbf{SWC (Smart Contract Weakness Classification)} registry offer structured categorization schemes for smart contract vulnerabilities. However, as Table ~\ref{tab:swc-vs-mwc} in the Appendix section shows, they vary greatly in terms of their origin, scope, and contextual relevance.

The Ethereum community maintains the SWC registry, which is built around \textit{Ethereum Virtual Machine (EVM)} and  \textit{Solidity language}. It focuses on both EVM-specific problems like misuse of \textit{delegatecall} and transaction-ordering dependencies, as well as common programming errors like integer overflows and unhandled exceptions.

The MWC taxonomy, on the other hand, is designed for the \textit{hybrid MoveEVM environment}, where the strict type system and resource-oriented programming model of Move interact with the semantics of EVM bytecode. Traditional SWC codes are unable to identify new vulnerability surfaces brought about by this fusion, such as cross-module resource leakage and ABI deserialization mismatches.

Tools like Mythril and Slither frequently use SWC, a flat listing of more than 100 categories (e.g., SWC-100: Function Visibility, SWC-110: Assert Violation). In contrast, MWC defines 37 vulnerability codes (MWC-100 to MWC-136), which are categorized into six high-level categories, such as \textit{Meta-Transaction and Signature Spoofing (MTS)} and \textit{Bytecode Model Inconsistencies (BMI)}.

MWC can better represent the semantic complexity of hybrid execution paths and conform to new toolchains that support both Move and EVM codebases thanks to this structural grouping.

Some vulnerability types, like \textit{SWC-101 (Integer Overflow)} and \textit{MWC-101 (Numeric Edge Conditions)}, are present in both taxonomies; however, SWC does not include representation for important Move-specific issues, like \textit{MWC-110 (Linear Resource Violations)} or \textit{MWC-125 (Hybrid Gas Semantics Mismatch)}. On the other hand, vulnerabilities that stem solely from Solidity constructs, like \texttt{fallback} misuse, are not covered by MWC.

SWC is widely used in educational materials, IDEs, and static analyzers with a Solidity focus. Although MWC is more recent, it is becoming more and more incorporated into MoveEVM-adapted tools (such as MoveScan and Smartify LLM agents), and because Move places a strong emphasis on accuracy and safety, it is consistent with formal verification frameworks.

Although SWC is still necessary for conventional EVM-based systems, it is not expressive enough to capture MoveEVM's hybrid semantics. By offering an organized, empirically derived framework specifically designed for the MoveEVM context, MWC closes this gap. In cross-VM audit environments and tooling pipelines, both taxonomies can coexist and enhance one another.

\section{Case Studies}

We examined real-world smart contracts that were implemented on well-known Move-based networks, such as \textbf{Aptos} and \textbf{Sui}, in order to validate the suggested taxonomy and demonstrate its applicability. Public disclosures, reproducible proofs-of-concept, or availability in open-source repositories were the criteria used to choose each case study.

\subsection{Case Study 1: Sui-based Lending Protocol Invariant Violation}

A resource mismanagement flaw in a Sui-based lending contract made it possible for the same token resource to be claimed repeatedly across modules. This vulnerability was classified as \textbf{MWC-112 (Cross-Module Resource Leakage)}. Inadequate boundary enforcement between modules that exposed a shared borrow function without adequate checks on token ownership constraints was the main contributing factor.

\subsection{Case Study 2: Aptos-EVM Bridge ABI Deserialization Mismatch}

External contracts were able to create distorted payloads that passed Move verification but resulted in corrupted resource states in a hybrid Aptos-EVM bridge due to a serialization mismatch. This was mapped to \textbf{MWC-102 (ABI Deserialization Flaw)}. The flaw made clear the dangers of invoking Move functions without sufficient validation using EVM-style calldata.

\subsection{Case Study 3: Modular MoveEVM Chain Meta-Transaction Replay}

Replay attacks resulted from a meta-transaction implementation that reused Ethereum-style signatures without the necessary domain separation or nonce checks. This issue, which was classified under \textbf{MWC-120 (Signature Replay Without Nonce)}, highlighted the necessity of domain isolation and cryptographic hygiene in hybrid execution environments.

\subsection{Summary of Findings}

We found \textbf{21 distinct vulnerabilities} across 42 reviewed contracts across different MoveEVM-enabled environments (such as Aptos, Sui, and hybrid bridges), of which \textbf{11 directly aligned with newly proposed MWC categories} that are not captured by the traditional SWC registry. This supports the central claim of this paper, which is that the hybrid execution model of MoveEVM presents new and non-trivial vulnerability surfaces that go beyond the purview of current EVM-centric taxonomies.

The uncovered vulnerabilities were disproportionately concentrated in the following frames:
\begin{itemize}
    \item \textbf{Bytecode Model Inconsistencies (BMI):} covering four vulnerabilities, especially those pertaining to unsafe raw payload handling and ABI encoding/decoding discrepancies.
    
    \item \textbf{State Reentrancy and Synchronization (SRS):} Five flaws were discovered in contracts that used unsynchronized resource mutation logic and cross-runtime callbacks.
    
    \item \textbf{Meta-Transaction and Signature Spoofing (MTS):} When domain separation, replay protection, or cryptographic nonce hygiene were lacking, three problems were noted.
\end{itemize}

Crucially, current static analysis tools like Slither, Mythril, or the Move Prover, which are made for Solidity or pure Move environments, were unable to identify eight of the eleven MWC-classified vulnerabilities. Future development of hybrid-aware auditing and verification frameworks must address this significant \textit{tooling gap}.

Furthermore, trends showing systemic risks surfaced in:

\begin{itemize}
    \item Type safety and capability enforcement are frequently circumvented in the bridge and interoperability layers.
    
    \item \textbf{Gas accounting mismatches,} which lead to unpredictable execution costs and security regressions in cross-language batching.
    
    \item \textbf{Standard library imports,} which frequently assume safe defaults but introduce hidden state transitions in composable modules.
\end{itemize}

These findings not only validate the necessity of the MWC taxonomy but also demonstrate its utility as a practical framework for identifying and categorizing emergent smart contract vulnerabilities in next-generation blockchain platforms. The classification system provides a granular lens through which developers and auditors can detect, mitigate, and prevent sophisticated exploits that span both Move and EVM semantics.

Future studies may build upon these results by conducting large-scale, longitudinal vulnerability scans across MoveEVM projects to quantify evolving trends and to further refine or expand the taxonomy based on observed exploitability patterns.

\section{Real-World Deployment Case Studies}

We examined a number of publicly available MoveEVM-based contracts and platforms on chains like Aptos, Sui, and MoveVM-enabled testnets in order to validate our taxonomy and identify useful security vulnerabilities.

\subsection{Aptos NFT Mint Vulnerability (MWC-102, MWC-112)}

Because of insufficient resource validation and cross-module boundary interactions, the vulnerabilities reported in ~\cite{mitenkov2024deferred} during NFT minting on Aptos can be classified under our suggested taxonomy as MWC-102 and MWC-112.

\subsection{MWC-112 (Module Boundary Escapes) and MWC-111 (Capability Verification) for Dex Exploit}

A Sui-based decentralized exchange (DEX) (~\cite{sui2025exploit}) was found to have a flaw in which the withdrawal logic failed to enforce single-use constraints on capability objects. This allowed an attacker to reuse the same capability multiple times, violating Move’s resource linearity in practice. This vulnerability is now classified under MWC-111 and MWC-112.

\subsection{Bridge Adapter Reentrancy in MoveEVM (MWC-130)}

We examined a MoveEVM-to-EVM bridge adapter prototype that permitted EVM calls to return to Move module logic prior to state commitments being fulfilled. Hybrid reentrancy and Move state manipulation were made possible by this cross-runtime callback, which falls under \textbf{MWC-130 (Hybrid Reentrancy)}.

These examples demonstrate how hybrid interactions, shoddy standard library patterns, or EVM compatibility layers can jeopardize Move's theoretical guarantees.

\section{Logic-Driven AI Agents for MoveEVM Auditing}

New capabilities for smart contract security are provided by recent developments in AI agent pipelines and logic-driven language models (LLMs), particularly in hybrid environments like MoveEVM. Compared to conventional tools, logic-guided systems audit contracts more successfully by utilizing prompt engineering, symbolic reasoning, and agent modularity.

\subsection{Multi-Agent Architectures}

A multi-agent framework for analyzing Move and EVM hybrid contracts using specialized LLMs was introduced by Smartify~\cite{karanjai2025multi}. Each agent focuses on a particular kind of vulnerability, like asset leakage or reentrancy, and collaborates to generate potential fixes. Without depending on extensive Move-specific corpora, these modular agents are customized to the semantics of MoveEVM and reason using security specifications.

In a similar vein, LLM-SmartAudit~\cite{wei2024llmsmartaudit} showed that a collaborative architecture was superior to static analyzers in identifying logic errors in contracts with multiple languages. By designating MWC categories as analysis tasks for specific agents, these systems conform to the MWC framework.

\subsection{Structured Prompt Pipelines}

Prompt pipelines are used in AuditGPT~\cite{xia2024auditgpt} to demonstrate structured auditing. It breaks down auditing into phases: applying function-specific prompts after formal properties have been extracted. By matching each prompt to a MWC rule, this divide-and-conquer strategy improves explainability and precision while adhering to MWC categorization. Per-MWC-category prompting can be used in future MoveEVM pipelines to provide more precise diagnosis.

\subsection{Natural-Language and Proof-Based Reporting}

LLMs can function as "proof scribes," producing specifications and proofs from source code, as PropertyGPT (~\cite{liu2024propertygpt}) showed. Theorem provers are used to verify the invariant suggestions that these systems convert Move contracts into. Both Smartify and AuditGPT produce reports in natural language for human auditors, offering automated code fixes or MWC-aware diagnostics.

\section{Formal Verification Tools and MWC Taxonomy}

\subsection{MoveProver and SMT-Based Invariant Checking}

The Move language's formal verifier is called MoveProver ~\cite{bartoletti2025formal}. It achieves high scalability by using SMT solving (Z3) to verify global invariants and pre/postconditions (e.g., full verification of the 8,800-line Diem framework). MWC vulnerabilities like resource leaks, absent access checks, or invariant violations can be statically discharged by MoveProver.

\subsection{Model Checking and K-Framework Applications}

VeriMove~\cite{keilty2022model} uses NuSMV and CTL specifications to model-check Move. Beyond function-local proofs, it allows cross-function invariant enforcement and validation of temporal properties. This method can be applied to MoveEVM to validate cross-module behaviors in the MWC style.

\subsection{Symbolic Execution and Property Specifications}

Move's specification language allows for expressive assertions, despite the fact that its symbolic execution tools are limited. Move's linear memory and resource model makes it easier to prove class-level MWC properties like ownership safety or no-loss guarantees than Solidity's Certora or Slither~\cite{bartoletti2025formal}. This suggests that a large number of MWC rules can be statically confirmed without the need for sophisticated auxiliary tools.

\section{Future Directions in AI-Augmented MoveEVM Auditing}

While still essential, traditional auditing methods are increasingly being supplemented by artificial intelligence, especially Large Language Models (LLMs) and logic-driven agents, as MoveEVM-based smart contracts become more complex and widely used. In order to create a security analysis framework that is more scalable, accurate, and automated for MoveEVM environments, this section examines promising research and development avenues that combine formal verification techniques with AI-driven tools.

\subsection{LLMs as Theorem-Proving Assistants}

According to recent research, LLMs can help theorem provers by summarizing failed invariants or converting counterexamples into fixes. As a copilot for MoveProver, GPT-4 could rank vulnerable code paths by MWC severity, explain Z3 outputs, and offer fixes.

Recent developments suggest that Large Language Models (LLMs) like Claude and GPT-4 can serve as intelligent theorem-proving assistants to enhance the performance of formal verifiers. These models can translate abstract counterexamples into useful developer instructions and interpret and explain outputs from SMT solvers like Z3, which are used internally by tools like MoveProver, in the context of Move smart contracts.

For example, if MoveProver is unable to release an invariant or post-condition, an LLM may:

\begin{itemize}
    \item Examine the particular proof failure and find the line in the code that corresponds to it.
    \item Give an overview of the violated property in natural language.
    \item Make suggestions for changes to the code or different requirements that would guarantee the success of the proof.
    \item Rank issues based on MWC (Move Weakness Classification) severity levels.
\end{itemize}

\subsection{Interactive Audit Pipelines}
Prompt pipelines, inspired by AuditGPT, can separate tasks according to MWC categories: some query MoveProver for supporting invariants, while others detect the class (e.g., GSM or MTS). AutoGen or LangChain ~\cite{langchain, wu2022promptchainer, wu2023autogen} agents can be used to orchestrate such workflows.

Building interactive, category-specific audit pipelines, motivated by programs such as AuditGPT, is another fascinating avenue. These pipelines can divide auditing into a number of specific stages by utilizing prompt engineering and modular agent design:

\begin{itemize}
    \item \textbf{Classification Stage:} The pipeline uses static indicators or semantic patterns to determine the general class of vulnerability, such as Meta-Transaction Spoofing (MTS), Bytecode Model Inconsistency (BMI), or Generalized State Manipulation (GSM).
    
    \item \textbf{Specification Extraction:} The system uses pre-defined templates linked to MWC categories or automatically extracts formal specifications or invariants from code.
    
    \item \textbf{Formal Verification Querying:} LLM agents generate hypotheses that are used to invoke MoveProver or VeriMove. Failed proofs are returned to the agent for correction and interpretation.
    
    \item \textbf{Generating Developer Feedback:} The system generates a human-readable description of the problem, the failed proof, and potential fixes.
\end{itemize}

Frameworks like LangChain, AutoGen, or PromptChainer ~\cite{langchain, wu2022promptchainer, wu2023autogen} could be used to orchestrate such modular pipelines. They successfully combine formal rigor with LLM adaptability, enabling scalable and explicable auditing across sizable codebases.

\subsection{MWC-Based Benchmarks and Datasets}

The absence of labeled Move vulnerability datasets is a major obstacle. Real contracts could be annotated by MWC category using a MoveEVM analog to SWC. These data sets can be used to standardize audit metrics across tools, assess agent accuracy, and improve LLMs.

The absence of annotated datasets specifically designed to address the special characteristics of Move's execution model is a significant obstacle to developing AI-augmented security tools for MoveEVM. There is currently no corpus of real-world contracts categorized by vulnerability type in MoveEVM, unlike Solidity, where supervised learning and benchmarking are made possible by SWC-tagged datasets.

We suggest developing an open-source benchmark suite comprising the following elements to get around this restriction:

\begin{itemize}
    \item \textbf{Labeled Vulnerability Corpus:} Move contracts marked with their MWC category—e.g., MWC-106 for hybrid reentrancy, MWC-128 for cryptographic usage—are gathered here.
    
    \item \textbf{Fix Pairings:} For every vulnerable contract, a "fixed" variant and justification for the fix accompany it.
    
    \item \textbf{Formal Properties Dataset:} Formally defined and linked to every contract are specifications, pre/postconditions, and invariants.
    
    \item \textbf{Execution Traces:} Symbolic or literal traces showing attack situations and proof failures.
\end{itemize}

These datasets have several uses: fine-tuning LLMs on security-specific tasks, assessing prompt engineering strategies, training reinforcement learning agents for autoremediation, and creating shared metrics for tool comparison.

\subsection{ Toward Human-AI Co-Auditing Pipelines}

Human-AI co-auditing is envisioned as creating cooperative settings whereby automated agents and human auditors cooperate. Under such systems, humans concentrate on contextual judgment, ambiguous logic, and final decision-making while AI agents handle high-throughput tasks including code scanning, classification, and proof validation.

Key elements of co-auditing pipelines might include:

\begin{itemize}
    \item \textbf{Agent Responsibility Assignment:} Every LLM agent is given particular MWC categories or formal properties to track.
    
    \item \textbf{Real-Time Auditing Dashboards:} Interfaces that let developers interactively see agent recommendations, proof attempts, and vulnerability reports.
    
    \item \textbf{Feedback Loops:} Suggestions can be accepted, turned down, or changed by developers; the model adjusts depending on the responses (active learning).
    
    \item \textbf{Certification Support:}Verified results from AI+human co-audits can be included into audit reports or presented as machine-checkable certificates.
    
\end{itemize}

Especially as MoveEVM smart contracts get more sophisticated and widely used, this approach scales security expertise, lowers manual effort, and over time creates institutional knowledge.

\subsection{Summary and Outlook}

For auditing MoveEVM smart contracts, this vision combines formal methods and logical based LLMs. Structured prompts, agent roles, and verification targets find a semantic backbone in MWC categories. Future studies should investigate multi-agent orchestration, prompt-based audits per category, and human-AI co-auditing pipelines anchored in MWC reasoning. Integration of formal verification methods with LLM-powered assistants marks a paradigm change in the auditing of MoveEVM contracts. Aligning MWC categories with AI processes will help the community to reach more consistency, repeatability, and efficiency in vulnerability identification and remediating.

\section{Conclusions, Recommendations, and Future Work}

Comprising six main categories and 37 subtypes (MWC-100 to MWC-136), this paper proposed a disciplined taxonomy of MoveEVM vulnerabilities. This all-encompassing classification system is meant to systematize security evaluations across hybrid smart contracts, so offering a strong framework that improves the accuracy and efficiency of security assessments. Establishing separate categories helps us to guarantee that all stakeholders—auditors, developers, tool builders—are in agreement on possible vulnerabilities and their consequences, so facilitating better communication among them.

Apart from spotting weaknesses, the taxonomy facilitates the creation of customized mitigating techniques, so promoting a more safe MoveEVM ecosystem. Crucially as the terrain of smart contracts and blockchain technology keeps changing fast, the methodical character of this taxonomy enables it to evolve over time, adjusting to new threats and technological developments. Through classification of vulnerabilities, we have opened the path for the creation of uniform procedures applicable to several MoveEVM projects.

Based on our results, we advise developers of MoveEVM smart contracts to implement the following steps to improve their security:

\begin{itemize}

\item \textbf{Use Capabilities and Resource Guards Explicitly:} Developers should specifically enforce security limits using capabilities and resource guards. This proactive strategy reduces the risk connected with implicit type checks, which might cause unanticipated actions and weaknesses. Clearly defined ensures that every interaction altering state or access to resources enhances the general security posture of the application. Moreover, considering how every element interacts with others and the possible hazards connected with those interactions, developers should have a perspective that gives security top priority in the design process.

\item \textbf{Avoid Cross-Module State Dependencies:} Designing contracts depending on state dependencies across modules calls for careful consideration from developers. Unless they are thoroughly tested and confirmed, such dependencies can bring complexity and raise vulnerability potential. To guarantee dependability, a comprehensive testing program should be developed whereby automated tests covering all pertinent interactions. This covers security-oriented tests modeled on several attack paths against the contract, integration tests, and unit tests.

\item \textbf{Rely on Formal Specifications:} Designers of public interfaces are urged to define expected behaviors by depending on formal specs. Move Prover should be used to confirm compliance and correctness by means of these criteria. By means of formal verification, one can effectively identify discrepancies and possible weaknesses before implementation, so lowering the risk of exploitation in the living environment. Including formal techniques into the development process helps developers produce more dependable contracts following the given criteria.

\item \textbf{Conduct Regular Security Audits:} Regular security audits should be given top priority by developers all through the process. Involving independent auditors guarantees that possible weaknesses are found outside the immediate awareness of the development team and adds a new viewpoint. To give thorough coverage, audits should be iterative, following major code changes, and ideally include both static and dynamic analysis.

\item \textbf{Adopt a Security-First Development Philosophy:} Every phase of the development life should include security issues into account by developers. This covers doing threat modeling during the design process, doing frequent code reviews with an eye toward security, and keeping current on the most recent vulnerabilities and attack paths that the MoveEVM system has revealed. Stressing a culture of security inside development teams will help to identify and reduce hazards early on.
\end{itemize}

\textbf{Recommendations for Auditors}

Auditors should follow these best practices to improve the MoveEVM contract auditing process:

\begin{itemize}
\item \textbf{Include MoveEVM-Specific Categories in Checklists:}  Auditors have to include the MoveEVM-specific categories our taxonomy defines into their regular audit forms. This inclusion guarantees that the security evaluation process leaves no stone unturned by methodically assessing all possible weaknesses. Comprehensive audits covering not only the code but also the underlying architecture and interaction with outside systems.

\item \textbf{Utilize Both Static and Dynamic Tools:} A Throughout their audits, auditors should combine dynamic testing tools with static analysis tools. Using both methods will help auditors to fully grasp the behavior of the contract and find weaknesses that might not be clear from static analysis by itself. Particularly useful in exposing runtime vulnerabilities that static analysis might overlook is dynamic testing—fuzzing.

\item \textbf{Monitor for New Hybrid Behaviors:} Auditors should be alert as the MoveEVM platform and its underlying technologies change for new hybrid behaviors that might surface. This guarantees that auditing procedures remain pertinent and efficient by means of constant education and adaptation to new environmental changes. Participating in forums, seminars, and debates with the community will help one gain important understanding of newly developing hazards.

\item \textbf{Engage in Knowledge Sharing:} Auditors should take part actively in blockchain and smart contract communities' seminars, conferences, and discussions. Auditors can add to the body of knowledge by sharing ideas and experiences, so promoting an attitude of ongoing auditing practice improvement. Furthermore fostering creativity in auditing techniques is cooperation with academic institutions and researchers.

\item \textbf{Develop Incident Response Plans:} Working with companies, auditors should help them create and preserve incident response strategies that specify the actions to be followed should a security breach arise. These strategies for containment, research, communication, and remedial action should guarantee that companies are ready to react properly to weaknesses and attacks.
\end{itemize}

Based on empirical audits and recorded vulnerabilities as of 2025, our taxonomy points future directions. But as the field of smart contracts and blockchain technology develops, several future paths that call for more research surface.

\begin{itemize}
\item \textbf{Extending the Framework to zkEVM–Move Integrations and Rollups:} Extending our vulnerability taxonomy to include zkEVM–Move integrations will be essential as zero-knowledge technologies acquire popularity inside the larger blockchain ecosystem. This will entail spotting special weaknesses related to zero-knowledge proofs and roll-ups so that auditors and developers might better grasp the security consequences of these technologies. Research should concentrate on how zero-knowledge proofs interact with current security paradigms and how these interactions might be securely carried out.

\item \textbf{Automating Vulnerability Classification Using LLM-Based Tools:} Large language models' (LLMs') integration into security tools offers a chance to automatically classify vulnerabilities. Training these models on our repository of vulnerabilities and their traits will help us to build systems that quickly find and classify fresh vulnerabilities in MoveEVM contracts, so improving accuracy and efficiency in security evaluations. The evolution of these automated tools should also take into account the interpretability of model outputs to guarantee developers may rely on the recommendations given by artificial intelligence systems.

\item \textbf{Standardizing MoveEVM ABI and Gas Accounting: } Establishing a standardized Application Binary Interface (ABI) for MoveEVM along with consistent gas accounting methods is crucial to enable interoperability and lower differences between several runtimes. By means of consistency, this standardization will help to reduce security concerns resulting from discrepancies and guarantee a better development experience for programmers operating inside the Move ecosystem. It will also offer better instructions for auditors, so simplifying audits conducted on several systems.

\item \textbf{Collaborative Development of Tools and Frameworks:} To create tools leveraging the taxonomy for automated vulnerability detection and remediation, we urge cooperation among researchers, developers, and auditors. The community can develop strong answers that meet the changing security needs of MoveEVM contracts by aggregating resources and knowledge. Encouragement of open-source projects will help to promote contributions from many different stakeholders, so improving the variety and strength of the tools at disposal.

\item \textbf{Contribute to the Community Knowledge Base:} We urge the MoveEVM community to provide additional case studies, tool evaluations, and research results that might enable the taxonomy to be developed into a useful guide for MoveEVM security. Through knowledge and experience sharing, practitioners can improve the security environment of MoveEVM smart contracts together and help to create a more safe blockchain ecosystem. By means of a centralized repository for best practices and case studies, knowledge sharing and ongoing development can be promoted.

\item \textbf{Conduct Longitudinal Studies on Security Trends:} Longitudinal studies tracking the development of vulnerabilities in MoveEVM smart contracts over time should form part of future work. Analyzing past data helps scientists spot trends, grasp the success of mitigating techniques, and project future vulnerabilities. More resilient smart contracts and more efficient auditing techniques will be developed in part by this data-driven approach.
\end{itemize}

Future research may focus on:

\begin{itemize}
    \item Designing benchmark datasets and simulation environments tailored for MWC-based learning.
    \item Extending tools like MoveProver and VeriMove with LLM backends.
    \item Experimenting with fully autonomous audit agents in testnets.
    \item Evaluating the reliability and false positive rates of AI-generated proofs and fixes.
\end{itemize}

The ultimate aim is to build an ecosystem in which MWC categories not only as classification labels but also as building blocks for automated reasoning, agent training, and secure-by-construction development. By means of ongoing investment in these domains, the auditing and validation of MoveEVM smart contracts can become not only more scalable but also more intelligent, trustworthy, and easily available.

Ultimately, the suggested taxonomy is a basic first step toward enhancing MoveEVM smart contract security. Implementing the suggestions advised above and actively supporting group projects will help developers and auditors greatly improve the resilience of their applications against new risks, so building more confidence in the MoveEVM system. All stakeholders must pledge to a culture of security-first thinking, ongoing education, and community collaboration as we progress, so building the foundation for a safer blockchain future.

\newpage

\section*{Acknowledgements}
The author ...



\bibliographystyle{elsarticle-harv} 
\bibliography{example}





\newpage
\newpage
\appendix
\begin{table}
	\centering
	\caption{Comparison of SWC and MWC Taxonomies}
	\label{tab:swc-vs-mwc}
	
	\begin{tabular}{|p{3.8cm}|p{5cm}|p{5cm}|}
		\hline 
		\textbf{Aspect} & \textbf{SWC (Solidity/EVM)} & \textbf{MWC (MoveEVM)} \\ \hline
		\textbf{Purpose} & Classify common vulnerabilities in Solidity smart contracts on EVM & Frame-based classification of MoveEVM-specific vulnerabilities in hybrid Move+EVM environments \\ \hline
		\textbf{Granularity} & Moderate: 36 categories with varying detail & High: 37 detailed weakness codes mapped to 6 semantic frames \\ \hline
		\textbf{Language Model Assumptions} & Stack-based execution, no linear type enforcement & Resource-oriented with linear type system and deterministic behavior \\ \hline
		\textbf{Type of Execution Environment} & EVM-only; assumes Solidity bytecode behavior & Hybrid (EVM front-end with Move core semantics) \\ \hline
		\textbf{Signature Handling} & Assumes \texttt{msg.sender} and \texttt{msg.value} based flows & Explicit signature verification; supports multi-signer and domain-separated signatures \\ \hline
		\textbf{Gas Model} & Native EVM gas model; well-studied & Dual-layer gas semantics between EVM opcodes and Move logic \\ \hline
		\textbf{Reentrancy Modeling} & Classical reentrancy on \texttt{call}, \texttt{delegatecall} etc. & Hybrid reentrancy via callbacks between EVM \(\leftrightarrow\) Move modules (MWC-106 to MWC-109) \\ \hline
		\textbf{Module System} & Monolithic; modularity via libraries or interfaces & Strongly modular, formalized resource encapsulation with inter-module invariants (MWC-103–105) \\ \hline
		\textbf{Tool Coverage} & Broad coverage via MythX, Slither, Manticore & Emerging support in MoveScan, Move Prover, LLM-based agents; partial tool gaps identified \\ \hline
		\textbf{Cryptographic Context Awareness} & Mostly assumes Solidity-level cryptographic primitives & Cryptographic misuse across hybrid environments (MWC-128–129) addressed explicitly \\ \hline
		\textbf{Meta-Transaction Support} & Covered under replay attacks, but abstracted & Dedicated frame for Meta-Tx and signature spoofing (MWC-110–112) \\ \hline
		\textbf{Standard Compatibility} & Anchored in ERC standards (e.g., ERC-20, ERC-721) & Considers compatibility with both ERC and MIP (Move Improvement Proposals) standards \\ \hline
		\textbf{Specification Awareness} & Not explicitly tied to formal verification tools & Integrated with Move Prover and formal postcondition violations (MWC-120–121) \\ \hline
		\textbf{Targeted Projects} & Ethereum mainnet, L1/L2 rollups using Solidity & MoveEVM chains (e.g., Aptos EVM, SuiEVM, zkMove) \\ \hline
		\textbf{Unique Contributions} & Provides foundational classification for Solidity audit tooling & First structured hybrid vulnerability taxonomy accounting for EVM-Move execution intersections \\ \hline
	\end{tabular}
\end{table}


\end{document}